# What is the Physical Explanation for the Very Large Ballistic Magnetoresitance Observed in Electrodeposited Nanocontacts?.

N. García

*Laboratorio de Física de Sistemas Pequeños y Nanotecnología.*

*Consejo Superior de Investigaciones Científicas, Serrano 144, Madrid*

*28006, Spain.*

(July 12, 2002)

Typeset using REVTEX

# Abstract

Recent experiments in approximately 10nm size electrodeposited Ni-Ni nanocontacts have shown ballistic magnetoresistance values of 700% stable during a week (Garcia et al Appl. Phys. Lett. 79, 4550(2001)) and very recently up to 3000% (Chopra and Hua, Phys. Rev. B 66,020403-1 (2002)). These values can provide very interesting magnetoelectronic devices integrated in the terabyte/square inch. Scattering in thin domain wall of the Ni-Ni nanocontacts, or quantized effects phenomena do not explain these values. An explanation is presented that requires the existence of a very thin, less than 1nm, dead magnetic layer grown during the electrodeposition process that is transparent to the electrons and has two roles: one, is to conserve spin in the electron conduction and the other, is to change and increase the polarization at Fermi level in the Ni.

Ballistic magnetoresistance [1–5] (BMR) is the next future solution to the problem of integration of magnetic memories in the terabyte/squared inch range. The very large values observed in BMR ( 700% by Garcia et al [4] and up to 3000%!! by Chopra and Hua [5]) can even change the logic in the read-write process due to the very large signal. Results of early experiments on BMR in Ni-Ni atomic contacts, smaller than 1 nm in size and having resistance larger than approximately 1000 Ohms were attributed by Tatara et al [6] to scattering by a sharp domain wall (DW), thinner than 1nm, following the theory of Cabrera and Falicov [7] Also, recently, Chung et al [8] have shown that the BMR of many different materials has a universal behaviour to DW scattering. Their calculations showed [6,8] that when the DW is thinner than 2nm the BMR is smaller than 10% because scattering of the DW do not conserve the spin in the conduction process.

However there have been recently experiments on BMR in Ni-Ni electrodeposited nanocontacts with resistances of a few Ohms, 10 to 30nm size, showing up to700-3000%!! BMR [4,5]. These very large values cannot be explained by the Ni-Ni DW scattering, be-cause calculations showed [9] that in this case the DW width is of the order of the contact size and then [6,8] the DW scattering provides 1% or smaller BMR values. This is very clear. Neither quantum conductance effects play any role in these later contacts because they have 1 Ohm resistance and then there are 10000 ! channels in the contact and at the RT values of the experiments quantum effects are simply ridiculously small.

In this short note I present a plausible explanation for the observed large BMR values in electrodeposited nanocontacts. The explanation is based on the assumption of a thin dead magnetic layer growing at some stage of the electrodeposition processes [4,9]. There are two requirements to be fulfilled to explain the large BMR values: i) spin has to be conserved during the electron conduction through the contact and ii) and more important, even if the spin is conserved, it should be clear that the polarization of Ni at Fermi level only can explain as much as 300% BMR [6–9] but not 3000%. This is the main problem. How can this happen?.

These two intriguing questions can be explained if one assumes that in the nanocontact a thin layer (smaller than 1nm) of dead magnetic material, say a complex of Ni with some other component in solution; is grown at some stage of the electrodeposition or is diffused to the contact from the bulk. This thin layer has to be transparent given the small resistance of the nanocontact (few Ohms). Then what happens is that this thin layer conserves spin during the conduction process because now there is not DW to scatter the electron and the process is completely non-adiabatic, no spin accommodation takes place from one side of the contact to the other in the antiparallel configuration (high resistance state) through the thin dead magnetic layer. Second is that the layer can change the Ni polarization at Fermi level at the interface of Ni and the thin dead layer. This change in polarization has to occur forcibly because there has to be charge transfer between Ni and the complex at the dead layer. It may be even possible to have an interface Ni-dead layer with full polarization for conduction electrons; i.e. to have an interface that is just an effective spin band. This could be obtained by having some O or S in the dead layer that form bonds with the sp electrons of Ni remaining also the the d electrons of Ni that at Fermi level are fully polarized. Then, the polarization will be unity and the BMR can grow to infinity!

The above proposal is consistent with, and explains the whole set of data on electrodeposited contacts:

1. The very large BMR values only happen in the electrodeposited samples. We have analyzed more that 10000 samples with nanocontacts grown by several different methods and we only observe very large values when the formation of the nanocontact is by electrodeposition. Including recent experiments that we have performed in nanocontacts grown by focussed ion beam, where we have to do electrodeposition to observe 500% BMR [10]

2. The fact that the BMR only depends on the particular contact for all the other parameters being fixed. We also find that it does not depend on the resistance (see ref.2 for a very clear illustration). This is in opposition to Chopra and Hua [5] who found a dependence with the contact size, or its resistance. We never found any such dependence.

3. Here is an important point, and this is that we and Chopra use the same electrolyte, a solution of NiSO4, and find large BMR results up to 1300%, published only up to 700% [4]. However Chopra and Hua [5]did an extra electrochemical treatment using KCl to prepare the tips, and, they obtained systematically larger values. This implies that the electrolyte is important in determining the kind of complex conforming the dead layer. This is also consistent with our explanation.

In conclusion, I have presented an explanation for the observed very large BMR values, that assumes the existence of a dead magnetic layer that conserves the electron spin in the conduction process and provides larger polarization at the Fermi level of the Ni-dead layer interface, needed to explain the large BMR values. It is now the challenge for scientists to investigate the nature of such layer as well as the means to conform it repetitively at will. If this is achieved, then BMR is without doubt the new effect that will give rise to a whole new set of magnetoelectronic devices integrated in the terabyte/square inch.

**Addition to the first version**: In recent discussions with J. W. Cahn has been also proposed that the grain boundary, where the nanocontact is formed, may pin a very thin DW and also change the density of states at Fermi level, involved in the electronic current, of the Ni at the grain boundary that will do the job required for the large observed BMR [11].

I thank Prof. J. A. Rausell- Colom for reviewing the manuscript. This work has been supported by the Spanish DGICyT.